\documentstyle[prl,aps,prb,epsfig,twocolumn]{revtex}
\newcommand{\be}{\begin{equation}}
\newcommand{\ee}{\end{equation}}
\newcommand{\bea}{\begin{eqnarray}}
\newcommand{\eea}{\end{eqnarray}}

\newcommand{\p}{\partial}

\newcommand{\la}{\langle}
\newcommand{\ra}{\rangle}

\newcommand{\ri}{\mbox{i}}

\def\nn{\nonumber\\}
\def\tp{$T_\perp$}

\def\r#1{(\ref{#1})}

\begin{document}
\draft
\twocolumn[\hsize\textwidth\columnwidth\hsize\csname %
 @twocolumnfalse\endcsname  
\title{Weakly coupled one-dimensional Mott insulators}

\author{Fabian H.L. Essler$^1$ and  Alexei M. Tsvelik$^{2}$}
\address{$^1$ Department of  Physics, University of Warwick,
Coventry CV4 7AL, UK\\
$^2$ Department of  Physics, Brookhaven National Laboratory, Upton, NY
11973-5000, USA} 
\maketitle

\begin{abstract}
We consider a model of one-dimensional Mott insulators coupled
by a weak interchain tunnelling $t_\perp$. We first determine the
single-particle Green's function of a single chain by exact
field-theoretical methods and then take the tunnelling into account by
means of a Random Phase Approximation (RPA). 
In order to embed this approximation into a well-defined expansion
with a small parameter, the Fourier transform $T_\perp(k)$ of the
interchain coupling is assumed to have a small support in momentum 
space such that every integration over transverse wave vector yields a
small factor $\kappa_0^2 \ll 1$. When \tp(0) exceeds a critical value,
a small Fermi surface develops in the form of electron and hole
pockets. We demonstrate that Luttinger's theorem holds both in the
insulating and in the metallic phases.
We find that the quasi-particle residue $Z$ increases very fast
through the transition and quickly reaches a value of about
$0.4-0.6$. The metallic state close to the transition 
retains many features of the one-dimensional system in the form of
strong incoherent continua.

\end{abstract}
]
\narrowtext
\sloppy

\section{Introduction}
The problem of the Mott metal-insulator transition, which is a consequence
of electron interactions and not due to the bandstructure, has been the
subject of great interest \cite{mottbooks} since the pioneering works
by Mott \cite{Mott}.
In 1964 Hubbard suggested a solution of this problem for a particular
model \cite{Hubbard}. The key point of
Hubbard's approach was to take into account the on-site Coulomb
interaction $U$ in zeroeth order and develop perturbation theory in $t/U$. Unfortunately, the transition
from the metallic to the insulating phase is expected to occur when the
tunnelling matrix element $t$ is comparable to the on-site repulsion $U$.
In this regime expansions in either $t/U$ or $U/t$ are not applicable.
To overcome this difficulty, various approaches have been suggested.
One of them considers a SU(N)-symmetric generalization of the
electronic model and carries out a $1/N$-expansion
\cite{Larkin,Kotliar88}. Another approach
is based on the so-called Dynamical Mean Field theory, which considers
a lattice in infinitely many dimensions $D\rightarrow\infty$ (see
\cite{Kotliar} and references therein and also \cite{Vollhard,florian}). 

In this paper we follow a different route. We consider a
quasi-one-dimensional model of interacting electrons, where the
tunneling along one direction is much larger than in all others. This
model can be described in terms of weakly coupled chains. When the
band is half-filled and the interchain tunneling is switched off, 
Umklapp processes dynamically generate a spectral gap $M$. The same
mechanism can generate gaps at {\sl any} commensurate filling
e.g. quarter-filling, but only if the interactions are sufficiently strong. 
In the half-filled case the decoupled chains develop a Mott-Hubbard
gap for any positive value of $U > 0$.  
In what follows we shall assume that the interchain tunneling is weak
but {\sl long-ranged}, such that the Fourier transform of  the
interchain tunneling matrix element has two strong peaks in the
Brillouin zone, one around zero wave vector and the other one around
${\bf Q} = (0, \pi, \pi)$ such that 
\bea
T_{\perp}({\bf k}) = - T_{\perp}({\bf k + Q})
\eea
Due to the long-range character of the tunneling, the width of these
peaks, denoted by $\kappa_0$, is small ($\kappa_0\ll 1$). Hence every
integration over the transverse momentum yields a small parameter
$\kappa_0^2$.  
The main idea of our approach is to treat the individual chains
non-perturbatively and then to employ the Random Phase Approximation
(RPA) to take into account the interchain tunneling. Since the
Mott metal-insulator transition is associated with development of
a coherent single-particle excitation branch, the correlation function
we are interested in is the single-electron Green's function $G$. The
RPA expression for $G$ is 
\bea
G(\omega,q, {\bf k}) = [G_0^{-1}(\omega,q) - 
T_{\perp}({\bf k})]^{-1}\ ,
\label{G2}
\eea
where $q$ is the momentum along the chain direction and $G_0$ is the
single-particle Green's function for an individual chain. 
The corrections to RPA are of higher order in $\kappa_0^2$, which is
the small parameter in our expansion.
The smallness of $\kappa_0^2$ suppresses multiparticle tunneling
processes, which generate exchange interactions between the chains and
eventually lead to a three-dimensional phase transition. However, 
for small $\kappa_0^2$ the temperature at which this transition occurs 
is much smaller than the characteristic energy scale $M$ of the
problem. Therefore throughout the paper we shall assume that we work
at temperatures much smaller than $M$ (such that the thermal effects
on the single-particle Green's function can be neglected) but much
larger than the transition temperature $T_c$, i.e. $T_c \ll T \ll M$.
We note that an analogous RPA has been used by Wen \cite{wen} to study
weakly coupled Luttinger liquids (LL), see also \cite{boies} for a
derivation based on functional integrals. An improved calculation for
the case of coupled LLs has recently been carried out by Arrigoni
\cite{arrigoni}. It is based on taking into account multipoint
correlation functions of upcoupled chains in a controlled way. In a
Mott insulator multipoint correlation functions are much more
difficult to determine than in a LL. We therefore postpone the
discussion of corrections to RPA until future publications. 

We also neglect the long distance tails of the Coulomb
interaction. This can be justified at finite temperatures if the
dielectric constant at small frequencies is large. 

The basic input of our  approach is $G_0$. To calculate this function we assume
that for an isolated chain the Mott-Hubbard gap $M$ is much smaller
than the band width $D_{\parallel}$. In this case one can use the
continuous field theoretical description in the low energy limit. The
corresponding field theory will described in detail in the next
section and is highly universal; all information about the underlying
lattice model is incorporated in just three dimensionless constants. 
Thus, our assumptions can be summarized by the inequalities
\bea
D_{\parallel} &\gg& M(U) \approx t_\perp(0) \gg t_\perp(|k| > \kappa_0) ,
\nonumber\\
T_c &\ll& T \ll M
\eea
where $(U)$ symbolizes dependence of the Mott-Hubbard gap from the
interactions. 
The smallness of $M$ does not imply that the interactions are weak.
For example, in the Hubbard model the field theoretical description
works well even for $U/t_{\parallel}$ as large as 2-3 (see, for
example, \cite{JGE}).
At such interaction strength one should expect the spin and charge
velocities to be considerably different. This is a consequences of
spin-charge separation, which is one of the most interesting features
of one-dimensional strongly correlated systems.
The effects of spin-charge separation in the insulating regime deserve
a special discussion, which will be given later in the paper.   

The outline of this paper is as follows. In section \ref{sec:model} we
introduce the model and discuss the limit of decoupled chains. In
section \ref{sec:coupled} we take the interchain coupling into account
and determine various physics properties.

In section \ref{sec:bechgaard} we discuss possible experimental
applications.
 
\section{The model}
\label{sec:model}
We take as a starting point the following lattice model
of correlated electrons
\bea
H&=&\sum_l H^{(l)} + \sum_{l,m,n,\sigma}t_{lm}\
{c^{(l)\dagger}_{n,\sigma}} c^{(m)}_{n,\sigma} +{\rm h.c.}\nn
H^{(l)}&=&-t\sum_{n,\sigma}
{c^{(l)\dagger}_{n,\sigma}}
c^{(l)}_{n+1,\sigma}+{\rm h.c.}
+U\sum_n n^{(l)}_{j,\uparrow}n^{(l)}_{j,\downarrow}\ .
\label{Hamiltonian}
\eea
Here $l,m$ label Hubbard chains and $n$ labels the sites along a
given chain. In the physically most interesting situation the
interchain hopping matrix elements $t_{lm}$ are taken to be equal to
$t_\perp$ if $l$ and $m$ are nearest neighbours and zero otherwise. 
For this choice of interchain hopping our calculational scheme is 
uncontrolled: there is no small expansion parameter. One may
nevertheless apply our scheme and hope that the corrections are small
in some regime of temperatures. On the other hand one may choose the
$t_{nm}$'s in such a way that a small expansion parameter is
introduced, as explained above and in more detail in Appendix
\ref{sec:RPA}.

Let us first consider the case of uncoupled chains
$t_\perp=0$. For weak repulsion $U< t$ and low energies a
field-theory description is appropriate. Keeping only modes in the
vicinity of the Fermi momenta $\pm k_F=\pm \pi/2a_0$, we may decompose
the lattice Fermi operators as
\be
c_{n,\sigma}^{(l)}=\sqrt{a_0}\left[\exp(ik_Fx)\ R^{(l)}_\sigma(x)+
\exp(-ik_Fx)\ L^{(l)}_\sigma(x)\right],
\ee
where $a_0$ is the lattice spacing. Inserting this prescription into
the Hamiltonian (\ref{Hamiltonian}) and dropping the chain index $(l)$
for the time being, one obtains 
\begin{eqnarray}
{\cal H}&=& \sum_{\sigma} v_F \int dx\left[L^{\dagger}_\sigma\ i\partial_x
L_\sigma - R^\dagger_\sigma\ i \partial_x R_\sigma\right]\nn
&&+\frac{g}{3}\int dx\left[:{\bf I}\cdot{\bf I}: + :{\bf
\bar{I}}\cdot {\bf\bar{I}}:
-:{\bf J}\cdot {\bf J}: - :{\bf \bar{J}}\cdot {\bf\bar{J}}: \right]\nn
&& +2g\int dx\left[ {\bf I}\cdot {\bf\bar{I}} - {\bf J}\cdot
{\bf\bar{J}}\right].
\label{hamil1}
\end{eqnarray}
where $v_F=2ta_0$ is the Fermi velocity and $g=U a_0$. Here $\bf J$
and ${\bf I}$ are the chiral components of SU(2) spin and pseudospin
currents
\begin{eqnarray}
I^3&=&\frac{1}{2}\sum_\sigma :L^\dagger_\sigma L_\sigma :\ ,\quad
I^+=L^\dagger_\uparrow L^\dagger_\downarrow\ ,\nonumber\\
\bar{I}^3&=&\frac{1}{2}\sum_\sigma :R^\dagger_\sigma R_\sigma :\ ,\quad
\bar{I}^+=R^\dagger_\uparrow R^\dagger_\downarrow\ ,\nonumber\\
{J}^3&=&\frac{1}{2}\left(L^\dagger_\uparrow L_\uparrow
-L^\dagger_\downarrow L_\downarrow\right)\ ,\quad
{J}^+=L^\dagger_\uparrow L_\downarrow\ ,\nonumber\\
\bar{J}^3&=&\frac{1}{2}\left(R^\dagger_\uparrow R_\uparrow
-R^\dagger_\downarrow R_\downarrow\right)\ ,\quad
\bar{J}^+=R^\dagger_\uparrow R_\downarrow\ .
\end{eqnarray}
By employing the Sugawara construction, the Hamiltonian (\ref{hamil1})
can now be split into two parts, corresponding to the spin and charge
sectors respectively \cite{GNT}
\begin{eqnarray}
{\cal H}&=& {\cal H}_c+{\cal H}_s\ ,\nonumber\\
{\cal H}_s&=& \frac{2\pi v_s}{3}\int dx\left[
:{\bf J}\cdot {\bf J}:+:\bar{\bf J}\cdot \bar{\bf J}:\right]
-2g\int dx\ {\bf J}\cdot \bar{\bf J}\ ,\nn
{\cal H}_c&=& \frac{2\pi v_c}{3}\int dx\left[
:{\bf I}\cdot {\bf I}:+:\bar{\bf I}\cdot \bar{\bf I}:\right]
+2g\int dx\ {\bf I}\cdot \bar{\bf I}.
\label{su2thi}
\end{eqnarray}
Here $v_s=v_F-Ua_0/2\pi$ and $v_c=v_F+Ua_0/2\pi$.
Apart from the (marginally) irrelevant current-current interaction in
the spin sector and the difference in spin and charge velocities, the
Hamiltonian (\ref{su2thi}) is identical to the one of the SU(2)
Thirring model. The SU(2) Thirring model can be bosonized in terms of
a Sine-Gordon model and a free boson; the resulting action density is
\bea
{\cal S}_s &=& \frac{1}{16\pi}\left[v_s^{-1}(\p_{\tau}\Phi_s)^2 +
v_s(\p_x\Phi_s)^2\right], \nn 
{\cal S}_c &=& \frac{1}{16\pi}[v_c^{-1}(\p_{\tau}\Phi_c)^2 +
v_c(\p_x\Phi_s)^2] +  \lambda\cos(\beta\Phi_c)\ ,
\label{SGM}
\eea
where $\beta=1$. If we consider additional small density-density
interactions between nearest neighbour sites in \r{Hamiltonian} the
field theory limit is again of the form \r{SGM}, but now $\beta<1$.
Using the integrability of the model \r{SGM} it is possible to
determine dynamical correlation functions. This is the subject of the
following section.

\section{Uncoupled chains}

The calculation of the spectral function for half-filled Mott
insulators is based on the following principles:

(a) \underline{Locality}: $R_{\alpha}, L_{\alpha}$ are local fields. 
This  allows us to employ the standard formfactor approach \cite{FF}. 
Some important elements of this approach are reviewed in Appendix
\ref{sec:FF}. 

(b) \underline{Spin-Charge separation}: The Hamiltonian (\ref{su2thi}) is
the sum of two parts representing the spin and charge degrees of
freedom respectively and by means of bosonization one can represent
$R_{\alpha}, L_{\alpha}$ as  products of fields  belonging to the
different sectors. For example: 
\bea
R_{\sigma} =\frac{\eta_\sigma}{2\pi}\ 
\exp\left(\frac{i}{2}\phi_c\right)\
\exp\left(\pm\frac{i}{2}\phi_s\right),
\label{R} 
\eea
where $\phi_c, \phi_s$ are chiral components of the bosonic
fields, $\eta_\sigma$ are Klein factors and the plus (minus) sign
corresponds to $\sigma=\uparrow$ ($\sigma=\downarrow$). The components
into which the creation and annihilation operators  are factorized are
nonlocal fields, but the factorization of the field means that all its
formfactors also factorize, at least in the limit when one of the
sectors becomes massless.  

(c) \underline{``Triviality'' of the spin sector}: In the limit $g_s
\rightarrow 0$ the field theory for the spin sector becomes massless
and the correlation function of the spin exponent in (\ref{R}) is
simply given by
\bea
\left\la\exp\left[\frac{i}{2}\phi_s(x,\tau)\right]\
\exp\left[-\frac{i}{2}\phi_s(0)\right]\right\ra 
= \frac{1}{\sqrt{v_s\tau - i x}}.
\label{spin} 
\eea

(d) \underline{Lorentz invariance and Watson's theorem}: Let us
consider formfactors of the left moving fermi operator
$L_\sigma(x)$. The first nonvanishing formfactor is between the vacuum
and a scattering state of one spinon (with spin $\sigma$) and one
antiholon. We denote the rapidity of the spinon (antiholon) by
$\theta_s$ ($\theta_c$) (our notations are summarised in Appendix
\ref{sec:FF}). Lorentz invariance and Watson's theorem impose the
following form for the first nonvanishing formfactor for the fermion
operator
\bea
\la 0|L_\sigma(0)|\theta_c,\theta_s\ra_{\bar{h}s} = \exp[(\theta_c +
\theta_s)/4]\ f(\theta_c - \theta_s)\ .
\label{matrix} 
\eea
The function $f(\theta)$ is periodic with period $2\ri\pi$ and does
not contain poles. Furthermore the usual asymptotic bound \cite{DM95}
yields
\be
\lim_{\theta_\alpha\to\infty}\la 0|L|\theta_c,\theta_s\ra\
\exp\left(-\frac{\theta_\alpha}{4}\right) \leq {\rm const}\ ,
\ee
where $\alpha=c,s$. This leaves us with the only possibility
$f(\theta) = \rm const$. As we see, this matrix element does not
depend on the anisotropy of the interaction (up to a constant
prefactor). The reason for this is that all information about the
anisotropy of the coupling constants is contained in the two-particle
S-matrices of holons and spinons and does not influence the
single-particle emission. An explicit expression for $f$ has
recently been obtained in \cite{LukZam01}; for $\beta^2 \rightarrow
1$ (the isotropic case) the numerical value is  
\be
f = (Z_0/2\pi)^{1/2}; ~~ Z_0=0.9218.
\ee

The matrix element (\ref{matrix}) describes the emission of one kink
in the charge and one kink in the spin sector. 
The charge part of this first term in the expansion is thus equal (up
to a overall numerical factor) to  
\bea
&&\int_{-\infty}^{\infty}d\theta\ e^{\theta/2}\exp[-
m\tau\cosh\theta - \ri m\frac{x}{v_c}\sinh\theta] \nn
&&= \left(\frac{\tau v_c - \ri
x}{\tau v_c + \ri x}\right)^{\frac{1}{4}}K_{\frac{1}{2}}(m\sqrt{\tau^2 +
x^2v_c^{-2}})\nn && =\frac{\exp[- m\sqrt{\tau^2 +
x^2v_c^{-2}}]}{\sqrt{v_c\tau + i x}} 
\label{holonG}  
\eea
In the next step we use the fact that the single-electron Green's
function factorizes into a charge and a spin piece. Let us consider
intermediate states containing a single antiholon and any number of
spinons. Factorization implies that if we carry out the sum over all
multi-spinon contributions, we must obtain the conformal result
\r{spin} for the spin sector. Therefore we arrive at the following
result for the two-point function
\bea
\la L_\sigma(x,\tau) L^\dagger_\sigma(0,0)\ra \simeq \frac{Z_0}{2\pi}\frac{\exp[-
m\sqrt{\tau^2 + x^2v_c^{-2}}]}{\sqrt{(v_s\tau +ix)(v_c\tau+ix)}}\ , 
\label{gl}
\eea
The leading corrections to \r{gl} involve intermediate states
containing two antiholons and one holon and are thus of order ${\cal
O}(\exp(-3mr))$. Similarly we have
\bea
\la R_\sigma(x,\tau) R^\dagger_\sigma(0,0)\ra \simeq
\frac{Z_0}{2\pi}\frac{\exp[- m\sqrt{\tau^2 +
x^2v_c^{-2}}]}{\sqrt{(v_s\tau-ix)(v_c\tau-ix)}}\ .
\label{gr} 
\eea

Eqns(\ref{gl}-\ref{gr}) were first written down by Wiegmann (for $v_c
= v_s$) \cite{vigman}. This remarkable result appears to have been
subsequently forgotten and rediscovered only much later by Voit, who
conjectured the same form the Green's function \cite{voit}. The
correct form was again reproduced by Starykh {\it et al.}
\cite{starykh}, although on the basis of arguments which cannot be
accepted as rigorous. An earlier attempt to calculate the single
particle Green's function using the formfactor  approach \cite{orgad}
led to an incorrect result yielding $K_1$-function  instead of
$K_{\frac{1}{2}}$ in Eq.(\ref{holonG}). Finally, Parola and Sorella
\cite{sorella} determined the spectral function by mapping the problem
of a single hole in a Mott insulator onto an effective spin chain with
skew boundary conditions. Their results are somewhat implicit so that
it is difficult to compare them to \r{gr}.

To conclude this section we would like to remark that the above
results can be generalized to the case of the SU(N) Thirring
model
\bea
&&\la L_\sigma(x,\tau) L^\dagger_\sigma(0,0)\ra = \frac{Z_0(N)}
{2\pi}\frac{1}{(v_s\tau + i x)^{\frac{1}{N}}}\nn
&&\times\left(\frac{v_c\tau - i x}
{v_c\tau + i x}\right)^{\frac{1}{2} - \frac{1}{2N}}K_{1 -
\frac{1}{N}}(m\sqrt{\tau^2 + x^2v_c^{-2}}) .
\eea
Here we have used that the kinks in the SU(N) Thirring model have spin
$\frac{1}{2}(1-1\frac{1}{N})$ \cite{sunthirring}. A similar formula
with $x \rightarrow -x$ holds for the right-moving fermions.  

\subsection{Short-distance behaviour: RG calculations.}

Strictly speaking, single kink emissions in the charge sector allow
one to consider only the frequencies up to $3m$ which is the threshold
for the emission of two antiholons and one holon. However, since the
numerical value of $Z_0$ is quite close to one eqns
(\ref{gl},\ref{gr}) are likely to provide a very good description
of the Green's function in the entire frequency interval. Indeed,
considering the short-distance (high energy) behaviour ($ m\sqrt{\tau^2 +
x^2v_c^{-2}} \ll 1$) in (\ref{gr}) and (\ref{gl}) we recover the
standard expressions for the Green's functions of the Luttinger liquid
(modulo the prefactor $Z_0$, which, as we have said, is almost 1). A
better description of the short-distance behaviour is obtained by
carrying out a renormalisation group (RG) analysis. The RG equations
for the coupling constant $g$ and wave function renormalisation $Z$ in the
case $v_c=v_s=v$ are 
\bea
g_0&=&g+\frac{g^2}{2\pi}\ln\frac{\Lambda^2}{\mu^2}+{\cal O}(g^3)\ ,\\
Z&=&1-\frac{3}{32\pi^2}g^2\ln\frac{\Lambda^2}{\mu^2}+{\cal O}(g^3)\ ,
\eea
where $\Lambda$ is a UV cutoff and $\mu$ a subtraction point.
The RG improved result for the two-point function is thus
\bea
\la R_\sigma(x,\tau) R^\dagger_\sigma(0,0)\ra =
\frac{Z(r)}{2\pi(v\tau-ix)}\ ,
\eea
where
\bea
Z(r)&\simeq&1-\frac{3}{16\pi}g(r)\ ,\nn
g(r)&\simeq&-\frac{\pi}{\ln(rm)}\ ,\quad r^2=x^2+v^2\tau^2\ll m^{-2}.
\eea
Comparing this to \r{gr} for short distances, we find
very good agreement.
In the anisotropic case, where the short-distance behaviour of the
Green's function is controlled by the Luttinger liquid exponent, the
agreement becomes worse. For ``large'' anisotropy it may become
necessary to take intermediate states with e.g. two solitons and one
antisoliton into account in order to get a good description of the
Green's function over a large range of energies.

\subsection{Retarded dynamical Green's function}

Given the expression \r{gr} for the Green's function in position space
we can calculate the dynamical Green's function by Fourier
transformation and subsequent analytic continuation to real
frequencies. A straightforward calculation gives the following result
for the retarded Green's function 
\bea
&&G^{(R)}_{RR}(\omega, q) =
-Z_0
\sqrt{\frac{2}{1+\alpha}}
\frac{\omega+v_cq}{\sqrt{m^2+v_c^2q^2-\omega^2}}\nn
&&\times
\left[\left(m+\sqrt{m^2+v_c^2q^2-\omega^2}\right)^2
-\frac{1-\alpha}{1+\alpha}(\omega+v_cq)^2\right]^{-\frac{1}{2}}\nn
\label{G}
\eea
where $\alpha = v_s/v_c$. At $\alpha = 1$ this simplifies to

\bea
G^{(R)}_{RR}(\omega, q) =
-\frac{Z_0}{\omega-vq}
\left[\frac{m}{\sqrt{m^2+v^2q^2-\omega^2}}-1\right].
\label{G1}
\eea
 
\subsection{Spectral function}

The spectral function is defined as 
\be
A_{RR}(\omega,q)=-\frac{1}{\pi}{\rm Im}G^{(R)}_{RR}(\omega,q)\ .
\ee

In the limit $v_c=v_s=v$ we obtain from (\ref{G1}) the following simple
result 
\bea
A_{RR}(\omega, q)=\frac{Z_0\ m}{\pi|\omega - vq|}
\frac{\Theta(|\omega|-\sqrt{m^2+v^2q^2})}{\sqrt{\omega^2 - m^2-v^2q^2}}\ .
\label{arr}
\eea
An explicit expression for the general case $v_s\neq v_c$ is easily
obtained from \r{G}. We note that the result is {\sl exact} for energies
$\omega\leq 3m$ as the intermediate states involving more than one
antiholon will start contributing only at this energy. However, from
previous experience \cite{CET} as well as on the basis of the
arguments given above, we expect \r{G} to be an excellent
approximation up to very high energies. 

For general $v_c,v_s$ the spectral function is nonvanishing in a
region with boundary $\omega= \epsilon(p)$ defined by the equation
\bea
\epsilon(p) = \mbox{min}_q[ v_s|p - q| + \sqrt{v_c^2q^2 + m^2}]\ .
\eea
The solution of this equation is 
\be
\epsilon(p)=
\cases{
\sqrt{m^2+v_c^2 p^2}\equiv\epsilon_0(p) & {\rm if}$\ p\leq Q$\cr
v_s p + m\sqrt{1 - \alpha^2}\equiv\epsilon_1(p) & {\rm if}$\ p\geq Q$\cr
}\ ,
\ee
where
\be
Q = \frac{m v_s}{\sqrt{1 - (v_s/v_c)^2}}\ .
\ee
The spectral function generally has several kinds of
singularities. Firstly, there are singularities just above the
threshold. For small momenta $|q|\ll Q$ there is a square root
singularity above the threshold $\omega=\epsilon(q)$
\bea
A(\omega,q)&\simeq& \frac{Z_0}{\pi}\frac{\epsilon_0(q)+v_cq}{\sqrt{2\epsilon_0(q)}}
\left([q+\epsilon_0(q)][\alpha\epsilon_0(q)-q]\right)^{-\frac{1}{2}}\nn
&&\times
\frac{1}{\sqrt{\omega-\epsilon_0(q)}}\ .
\eea
There also is a square root singularity above the threshold for
``large'' momenta $|q|\gg Q$, where
\bea
A(\omega,q) \simeq \frac{Z_0}{\pi\sqrt{v_c- v_s}}
\frac{1}{\sqrt{|q|-Q}}\frac{1}{\sqrt{\omega-\epsilon_1(q)}}\ .
\eea
This square root singularity is a general feature of the half-filled
Mott insulator and apparently holds for any ratio of the gap to the
bandwidth. Calculations done for a model where this ratio is
infinite \cite{laughlin} gave the same singularity.

\begin{figure}[ht]
\begin{center}
\epsfxsize=0.4\textwidth
\epsfbox{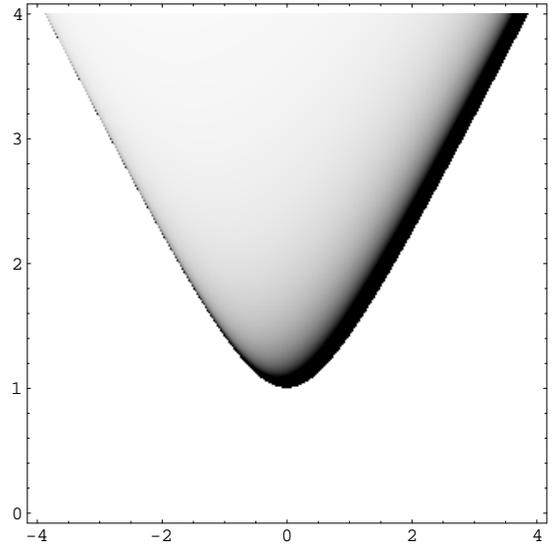}
\end{center}
\caption{\label{fig:aeq1}
Density plot of the spectral function $A_{RR}(\omega,q)$ as a function
of $\omega$ and $vq/m$ for $v_s=v_c=v$.
}
\end{figure}

\begin{figure}[ht]
\begin{center}
\epsfxsize=0.45\textwidth
\epsfbox{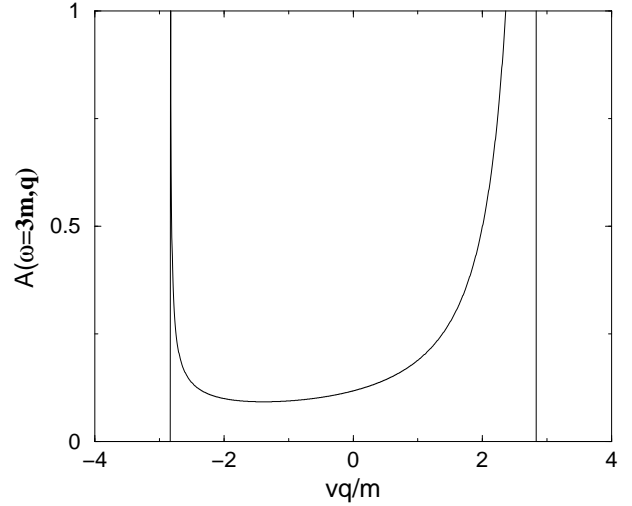}
\end{center}
\caption{\label{fig:a1o3}
Constant energy ($\omega=3m$) scan of the spectral function for $\alpha=1$.
}
\end{figure}

\begin{figure}[ht]
\begin{center}
\epsfxsize=0.4\textwidth
\epsfbox{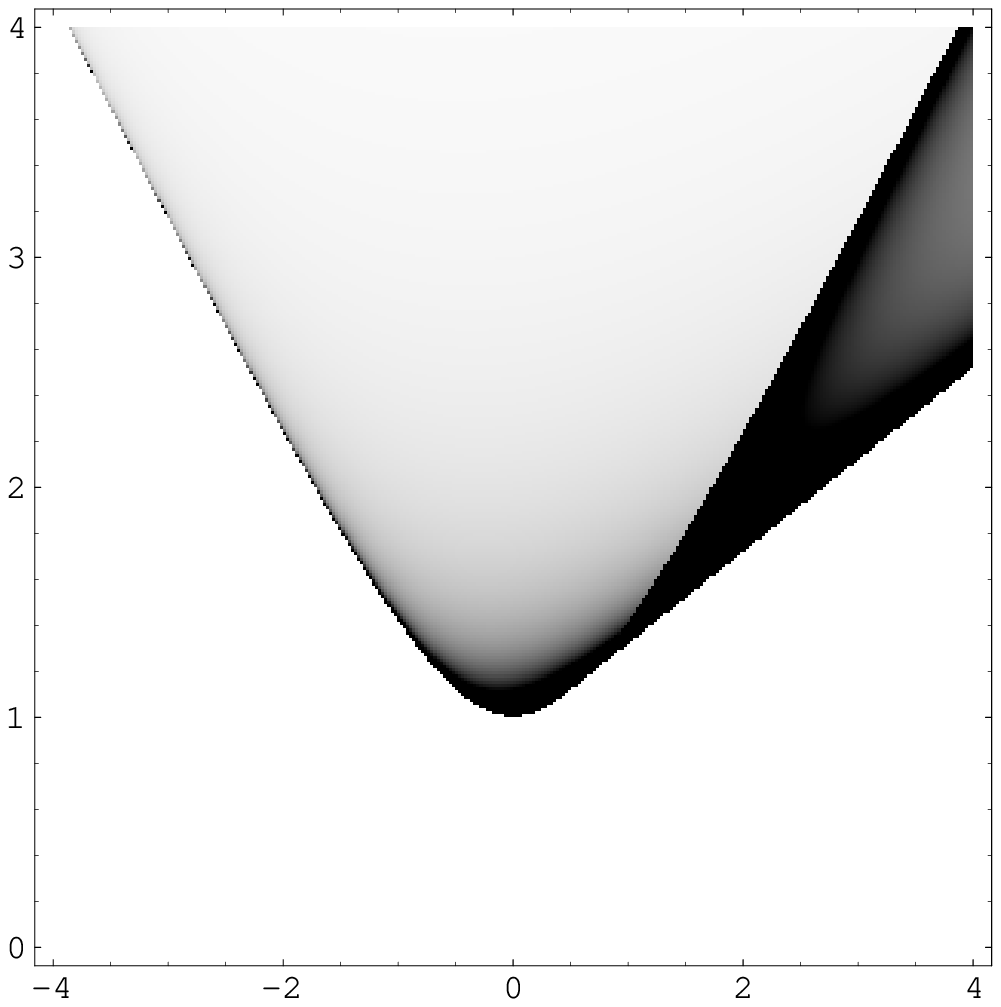}
\end{center}
\caption{\label{fig:aeq04}
Density plot of the spectral function $A_{RR}(\omega,q)$ as a function
of $\omega$ and $v_cq/m$ for $\alpha=0.4$.
}
\end{figure}

\begin{figure}[ht]
\begin{center}
\epsfxsize=0.45\textwidth
\epsfbox{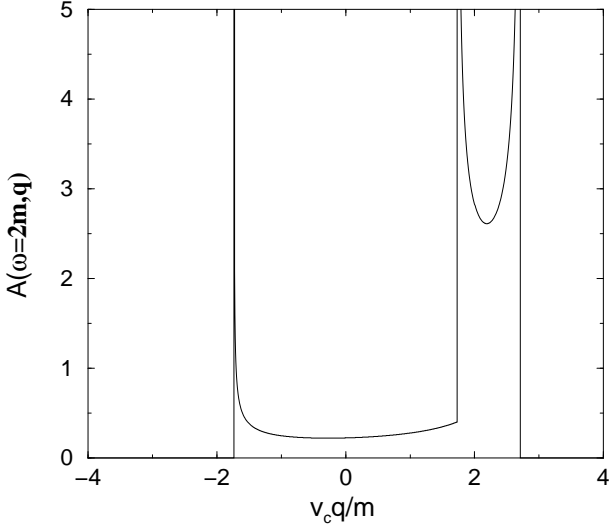}
\end{center}
\caption{\label{fig:a04o2}
Constant energy ($\omega=2m$) 
scan of the spectral function for $\alpha=0.4$.
}
\end{figure}

In addition there is a second square-root singularity for
$\omega\to\epsilon_0(q)$ from below
\bea
A(\omega,q)&\simeq& \frac{Z_0}{\pi}\frac{\epsilon_0(q)+v_cq}{\sqrt{2\epsilon_0(q)}}
\left([q+\epsilon_0(q)][v_cq-\alpha\epsilon_0(q)]\right)^{-\frac{1}{2}}\nn
&&\times
\frac{1}{\sqrt{\epsilon_0(q)-\omega}}\ .
\eea
For $q=Q$ these two singularities collapse onto a single one
and we have
\bea
A(\omega, Q) \simeq
\frac{Z_0}{\pi}\left(\frac{1+\alpha}{1-\alpha}\right)^\frac{1}{4}
2^{-\frac{5}{4}}
\epsilon(Q)^{-\frac{1}{4}}(\omega-\epsilon(Q))^{-\frac{3}{4}}\!.
\label{appr1}
\eea
In Fig.\ref{fig:aeq1} we show the spectral function for $v_c=v_s=v$ as
a function of $\omega$ and $q$. As $Q=\infty$ in this case the
threshold is simply given by $\omega=\sqrt{m^2+v^2q^2}$. The square root
singularities above the threshold are visible as regions of high
spectral weight. 
Fig.\ref{fig:a1o3} shows a constant energy scan at $\omega=3m$. 
When we allow the spin and charge velocities to be different an
interesting effect occurs as can be seen from the density plot of
$A_{RR}(\omega,q)$ in Fig.\ref{fig:aeq04}. For momenta $q<Q$ the
threshold occurs at the gap for a single soliton moving with momentum
$q$ and the spectral function looks very much like it did in the case
$\alpha=1$. However, at momenta $q>Q$ the threshold gets shifted
to $\omega=v_sq+m\sqrt{1-\alpha^2}$ and a lot of spectral weight is
concentrated between this threshold and the line
$\omega=\sqrt{m^2+v_c^2q^2}$, where a second singularity occurs (see
above). This is shown in the constant energy scan Fig.\ref{fig:a04o2}.
For large momenta $v_{s,c}q\gg m$ the double singularity at
$\omega\simeq v_{c,s}q$ is similar to what occurs in a Luttinger
liquid \cite{LL}. 

The states located in the ``wedge'' $\epsilon_1(q) <\omega
<\epsilon_0(q)$ correspond to a situation where the single antiholon
carries only a small part of the total momentum and the remainder is
made up by exciting (many) spinons. This effect is reminiscent of
Cherenkov radiation produced by a heavy particle moving through a
medium with a speed greater than the speed of light in this medium. 

\subsection{Tunneling density of states} 

From the Matsubara Green's function at coinciding space points in
model (\ref{su2thi}) we extract the single-particle density of states
\bea
\rho(\omega) = -2{\rm Im}\ G^{(R)}(\omega,x=0)=\frac{
2Z_0}{v_sv_c}\theta(|\omega| - m)\ .
\label{dos}
\eea
The density of states \r{dos} is {\sl constant} for energies
higher than the single-particle gap $m$. Formula \r{dos} is 
exact for $|\omega| \leq 3m$ and expected to be an excellent
approximation up to very high energies.

\section{Coherent single-particle mode generated by the interchain
tunnelling} 

Taking the interchain tunneling into account in the RPA yields an
expression for the single-electron function in the form of (\ref{G2}). 
The interchain tunneling gives rise to a branch of coherent
excitations below the threshold of the 1D spectral function $A(\omega,
q)$ in the region where $T_{\perp}(\vec{k}) < 0$. Therefore, even an
infinitesimal interchain coupling leads to the coherent particle
motion in the transverse direction and there is no confinement in the
sense of Anderson \cite{anderson}.

\subsection{$v_s=v_c$}
Let us first discuss the case $v_s=v_c=v$. For very small interchain
tunnelling $|T_{\perp}(\vec{k})| \ll m$ the pole of (\ref{G2})
appears at a finite frequency very close to $\epsilon_0(p) = \sqrt{p^2
+ m^2}$. In the vicinity of the pole we have  
\bea
G(\omega,q, \vec{k}) \approx \frac{Z(q,\vec{k})}{\omega -
\epsilon(q,\vec{k})}\ , 
\eea
where $Z(q,\vec{k})$ varies strongly at small $q$ and where for
$|vq|\ll |m^2/Z_0T_\perp(\vec{k})|$ 
\bea
\epsilon(q,\vec{k}) = \epsilon_0(q)-\left(\frac{Z_0T_\perp(\vec{k})m}
{\sqrt{2\epsilon_0(q)}[\epsilon_0(q)-vq]}\right)^2\ .
\eea
In the chain direction, the dispersion of the coherent mode is
asymmetric around $q=0$ and has a minimum at $vq\simeq
[Z_0T_\perp(\vec{k})]^2/m$. 
For larger values of $T_\perp(\vec{k})$ this picture remains
qualitatively unchanged although the dispersion law becomes more
complicated. In the vicinity of the minimum the dispersion is
approximately given by
\bea
\epsilon(q,\vec{k}) \approx\sqrt{\omega_0^2+v^2(q-q_0)^2}\ ,
\eea
where $\omega_0$, $q_0$  depend only on $T_\perp(\vec{k})$.
At large positive $vq\gg m$ we have
\be
\epsilon(q,\vec{k}) \approx 
vq+Z_0T_\perp(\vec{k})\ ,
\ee
and at large negative $vq\ll -\max(m,|Z_0T_\perp(\vec{k})|)$
\be
\epsilon(q,\vec{k})
\approx\sqrt{m^2+(vq)^2-\bigl[\frac{Z_0T_\perp(\vec{k})}{2vq}\bigr]^2}\ .
\ee

\begin{figure}[ht]
\begin{center}
\epsfxsize=0.45\textwidth
\epsfbox{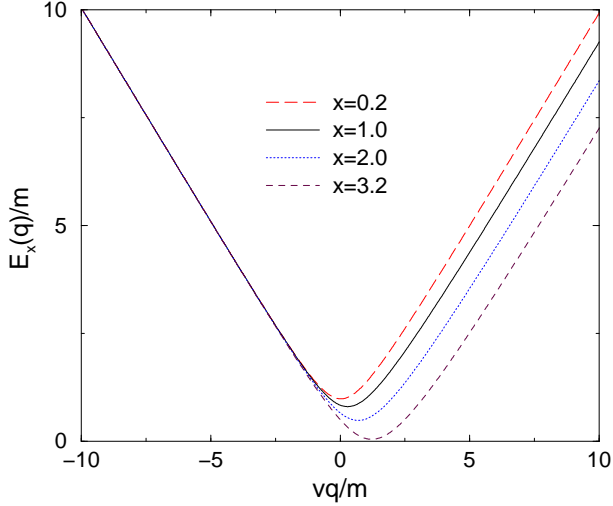}
\end{center}
\caption{\label{fig:cmode}
Dispersion of the coherent mode along the chain direction for several
values of $x=-Z_0T_\perp(\vec{k})$ and $\alpha=1$.
}
\end{figure}
The dispersion of the coherent mode depends on the transverse momentum
only through $T_\perp(\vec{k})$. In Fig.\ref{fig:cmode} we plot 
\be
E_x(q)\equiv \epsilon(q,\vec{k})\bigg|_{-Z_0T_\perp(\vec{k})=x}
\ee
as a function of $q$ for several values of $x$. In Fig. 6 we plot the
residue of the coherent mode for various $x$ versus $qv/m$.  
\begin{figure}[ht]
\begin{center}
\epsfxsize=0.45\textwidth
\epsfbox{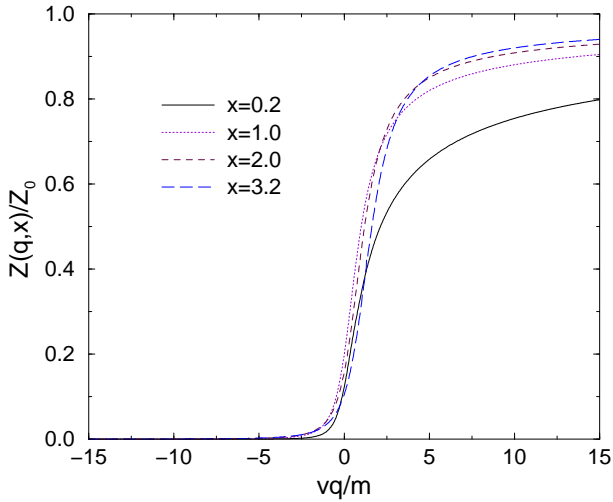}
\end{center}
\caption{\label{fig:zweight}
Residue of the coherent mode as a function of the momentum along the
chain direction for several values of $x=-Z_0T_\perp(\vec{k})$ and
$\alpha=1$. 
}
\end{figure}

We see that $Z(q,\vec{k})$ is very small within the noninteracting
Fermi surface.

\subsection{$v_s\neq v_c$}
\begin{figure}[ht]
\begin{center}
\epsfxsize=0.4\textwidth
\epsfbox{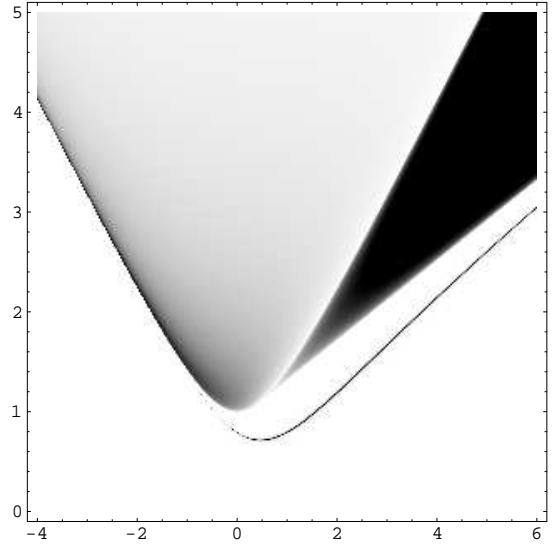}
\end{center}
\caption{\label{fig:al04t05}
Density plot of the spectral function $A_{RR}(\omega,q,\vec{k})$ as a
function of $\omega$ and momentum transfer along the chain direction
$q$ for $Z_0T_\perp(\vec{k})=0.5$.
}
\end{figure}

The case of different velocities can be dealt with in a completely
analogous way. In Fig.\ref{fig:al04t05} we show the spectral function
as a function of energy and momentum transfer along the chain
direction for weak tunnelling $Z_0T_\perp(\vec{k})=0.5$. The coherent
mode is clearly visible below the threshold of the one dimensional
spectral function.

\begin{figure}[ht]
\begin{center}
\epsfxsize=0.4\textwidth
\epsfbox{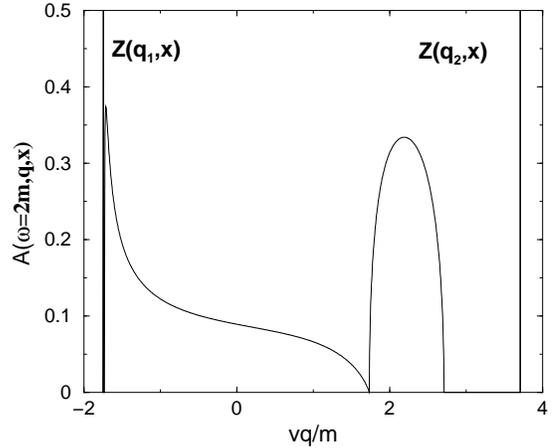}
\end{center}
\caption{\label{fig:a04o2t05}
$A_{RR}(\omega,q,\vec{k})$ as a function of the momentum along the
chain direction for $\omega=2m$, $\alpha=0.4$ and
$Z_0T_\perp(\vec{k})=0.5$. 
}
\end{figure}

In Fig.\ref{fig:a04o2t05} we show a constant energy scan of the
spectral function.

If we compare Fig.\ref{fig:a04o2t05} to the corresponding plot of the
spectral function of uncoupled chains in Fig.\ref{fig:a04o2}, we
notice that now there are two delta function peaks corresponding to the
coherent mode, and that the singularities of the incoherent scattering
continuum have been smoothed out. It turns out that most ($\simeq
75\%$) of the spectral weight {\sl for fixed} $T_\perp({\bf k})$
(i.e. we integrate only over $q$) is located in the coherent mode at
$q_2$, about $23\%$ sits in the incoherent spinon-antiholon continuum
and only $2\%$ is due to the coherent branch at $q_1$. The sharp
difference between the weights in the coherent modes is consistent
with the picture presented on Fig. \ref{fig:zweight}. 

As $t_\perp$ increases more and more of the spectral weight
gets transferred to the coherent mode. For example, for
$Z_0T_\perp(\vec{k})=3.2m$ approximately $95\%$ of the spectral weight
at energy $\omega=2m$ is located in the coherent mode at $q_2$. 
However, it is important to note that the coherent modes dominate only
a small portion of the Brillouin zone (the part where $T_\perp({\bf
k})$ is large). If we consider the total spectral weight,
i.e. integrate over the transverse momentum ${\bf k}$ as well, we find
that the contribution of the coherent mode is generally small.

\section{Fermi surface and Luttinger's theorem}

According to Luttinger's theorem the total number of particles in
the system is proportional to the volume of momentum space included in
the surface defined by the singularities of $\ln G(\omega =0, {\bf
p})$. These singularities may be due to either poles or {\it zeroes}
of the single-particle Green's function. As was recently emphasized by
Dzyaloshinskii \cite{Dz}, the latter possibility must occur in Mott
insulators thus removing an apparent violation of Luttinger's
theorem by the interactions. Indeed the Green's function (\ref{G})
exhibits precisely this property: at $\omega = 0$ it vanishes at $q
= 0$, which corresponds to the Fermi surface of the non-interacting
system. Since the position of the zeroes is unchanged in RPA, our
results are in agreement with Luttinger's theorem.   

When $t_{\perp}$ exceeds a critical value, electron- and hole-like 
pockets of Fermi surface appear.   
The Fermi surface is determined by the equation
\bea
G_0^{-1}(0, q) = T_{\perp}({\bf k})
\eea
The volumes of electron and hole pockets are equal since
$T_{\perp}({\bf k}) = -  T_{\perp}({\bf Q+k})$ where $Q =
(\pi,\pi,0)$. So  Luttinger's theorem continues to hold.

Using the fact that at $\omega = 0$ the Green's function is
always real, we get from (\ref{G}) 
\bea
[Z_0T_{\perp}({\bf k})]^2 &=& \frac{(v_c^2q^2 + m^2)(m + \alpha\sqrt{v_c^2q^2 +
m^2})}{\sqrt{v_c^2q^2 + m^2} - m},\nn
\mbox{sign}(T_{\perp}({\bf k}))&=&  -\mbox{sign}(q). 
\label{FS}
\eea
The critical value of $T_{\perp}({\bf k})$ necessary to produce the
solution is 
\bea
\left[\frac{Z_0t_{\perp}^{\rm min}}{m}\right]^2 =
3+\frac{9\alpha+1}{2}x_0\ ,
\eea
where and $x_0$ is determined by the momentum $q_0$ where the Fermi
surface first appears
\bea
x_0=\sqrt{1+v_c^2q_0^2/ m^2}=
\frac{3\alpha - 1 + \sqrt{1 + 10\alpha + 9\alpha^2}}{4\alpha}\ .
\eea
The critical value of $Z_0 t_{\perp}^{\rm min}$ varies from $2m$ at
small $v_s$ to $\sim 3.3 m$ for $v_s = v_c$. 

The residue at the Fermi surface is given by 
\bea
Z = \frac{2Z_0}{(1 + \alpha)}\frac{(x -1)}{x}\left[\frac{(x\alpha +
1)}{(x + 1)}\right]^{1/2},
\eea
where $x \equiv \sqrt{1+v_c^2q^2/m^2}$. Near the critical value of
$T_{\perp}({\bf k})$ the residue is numerically small but never goes
to zero. For example, in the limit $v_s \rightarrow 0$ when $x = 2$ we
get
\[
Z(v_s \rightarrow 0) \approx 0.58\ Z_0
\]
and at $v_s = v_c$ when $x \approx 1.62$ we have
\[
Z(v_s = v_c) \approx 0.38\ Z_0.
\] 
For a cubic lattice the Fermi surface forms an {\sl electron pocket}
around $\tilde{q}=q_0>0$, $k_0^y=k_0^z=0$ and {\sl hole pockets}
around $\tilde{q}=-q_0$ and $k_0^y=\pm\pi$, $k_0^z=\pm\pi$. The
volume of the electron pocket is the same as the sum of the
volumes of the hole pockets. When the pockets are very small their
shape can be determined by expanding Eq.(\ref{FS}) around the point
$({\bf 0},q_0)$, using that  
\[
T_{\perp}({\bf k}) \approx  T_{\perp}({\bf 0})  [1- {\bf
k}^2_{\perp}\gamma^2]\ . 
\]
Here ${\bf k}_\perp$ denotes the deviation in the transverse
direction from ${\bf 0}$ and $\gamma$ is proportional to the lattice
spacing in the transverse directions $a_\perp$. We obtain
\bea
&&2 T_{\perp}^2({\bf 0})\gamma^2{\bf
k}_\perp^2+ \frac{b}{2}(q- q_0)^2=T_\perp({\bf 0})^2-(t_\perp^{\rm
min})^2\ , 
\label{fermisurface}
\eea
where 
\[
b =\frac{v_c^2}{Z_0^2} \frac{(x_0 + 1)(4\alpha x_0 - 3\alpha
+1)}{x_0(x_0 - 1)}\ .
\]

We also can estimate the anisotropy of the Fermi surface from
(\ref{fermisurface}). The anisotropy of the masses is given by
\bea
\frac{m_{\perp}}{m_{\parallel}} = \frac{v_c^2}{4\gamma^2m^2}\frac{(x_0 +
1)(4\alpha x_0 - 3\alpha +1)}{x_0^3(1+\alpha x_0)}\ ,
\eea
where we have further approximated $T_\perp({\bf 0})\approx
t_\perp^{\rm min}$. Using that $v_c\simeq 2ta_{\parallel}$, where
$a_{\parallel}$ is the lattice spacing along the chains, we find that 
\be
\frac{m_{\perp}}{m_{\parallel}} \simeq {\cal A}\
\frac{a_\parallel^2}{\gamma^2} 
\ \frac{t^2}{m^2} , 
\ee
where ${\cal A}$ varies between $1.06$ for $v_s\to v_c$ and $0.62$
for $v_s\to 0$. 
Thus the magnitude of the mass ratio is determined by the competition
of two factors one of which is large ($t/m$) and the other is small
($a_\parallel/\gamma$).  
As a result, the Fermi surface may not be very anisotropic.

An obvious question is whether or not the RPA can be trusted to
describe correctly the formation of a Fermi surface. An obvious
problem of the RPA is that it automatically reproduces a purely one
dimensional result at the particular wave numbers ${\bf p}$ where
the transverse hopping vanishes $T_\perp({\bf p})=0$. On a 2D square
lattice this would be at the points $p_y=\pm\pi/2$.
In the case of coupled Luttinger liquids the improved treatment of
\cite{arrigoni} indicates that in the vicinity of these points RPA is
unreliable. In the case of coupled Mott insulators, the electron and
hole pockets we find for sufficiently large $t_\perp$ are located at
positions far away from the points where $T_\perp({\bf k})$ vanishes.

Recently a dynamical mean field theory approach has been developed, which
replaces the quasi-1D system by a single effective chain, from which
electrons can hop to a self-consistent bath
\cite{Georges01,biermann01}. The resulting model has to be analyzed by
numerical methods and for coupled Hubbard chains it is found that for
sufficiently large transverse hopping amplitudes an open Fermi surface
(close to the one of the noninteracting model) is formed. This is in
contrast to the electron and hole pockets we find in the RPA.

\section{Transverse Conductivity and Density of States}
\label{sec:coupled}

Using the results for uncoupled chains as well as the RPA
expression for the Green's function of weakly coupled chains we can
determine various other physical quantities. 

\subsection{The transverse conductivity}
Let us consider a situation where the transverse hopping is only
between nearest neighbour chains. The transverse current operator is
then given by
\be
j_\perp(x,l)=iet_\perp\left[R^{(l)}_\sigma(x){R^{(l+1)}_\sigma}^\dagger(x)
-{\rm h.c.}+ R\to L
\right] ,
\ee
where $l$ is a chain index and $x$ denotes the position along the
chain direction.
Using this expression we can analytically determine the leading
contribution in $t_\perp$ to the transverse conductivity by using the
result for the Green's function of uncoupled chains. We find
\bea
\sigma_\perp(\omega)&=&\frac{2Z_0^2e^2t_\perp^2}{\pi(v_c-v_s)}
\frac{1}{\omega}
\arctan\frac{4m\delta\sqrt{\omega^2-4m^2}}{\omega^2+\delta(\omega^2-8m^2)}\ ,
\eea
where $\delta=(v_c-v_s)/(v_c+v_s)$. 

In the limit $v_s\to v_c=v$ this simplifies to
\bea
\sigma_{\perp}(\omega) = \frac{2Z_0^2e^2t_\perp^2}{\pi
v}\frac{1}{\omega} (2m/\omega)^2\sqrt{(\omega/2 m)^2 - 1} 
\label{trans}
\eea

We see that the transverse conductivity vanishes at the threshold
$\omega=2m$ and increases above it in a characteristic square root
fashion. This is reminiscent of the behaviour found for the
longitudinal conductivity in \cite{CET}.

\subsection{Density of States}
\begin{figure}[ht]
\begin{center}
\epsfxsize=0.4\textwidth
\epsfbox{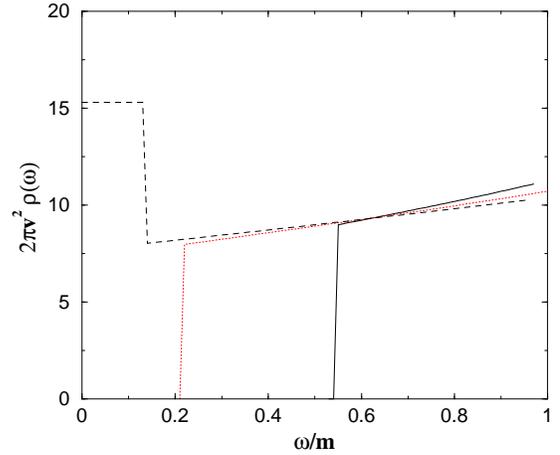}
\end{center}
\caption{\label{fig:dos2D}
Density of states for a 2D square lattice for frequencies smaller 
than the Mott gap. The curves are for $t_\perp=2$ (solid), $t_\perp=3$
(dotted) and $t_\perp=4$ (dashed).
}
\end{figure}

Within the RPA we can determine the density of states by integrating
the RPA spectral function over all momenta. This needs to be done
numerically. For simplicity we only consider the case $v_c=v_s=v$.
For a 2D system with $T_\perp(k_y)=t_\perp\ \cos(k_y)$ \cite{comment1}
we obtain the
results shown in Fig.\ref{fig:dos2D}. As $t_\perp$ is increased, the
Mott gap in the DOS is filled in and eventually a peak forms around
zero energy. 
The analogous analysis for a quasi 3D system with
$T_\perp(k_y,k_z) = t_\perp \left[\cos(k_y) + \cos(k_z) \right]$ yields a
qualitatively different result. As $t_\perp$ is increased the Mott gap
is again filled in, but now the DOS around $\omega=0$ remains
basically flat and no peak develops in the regime where RPA applies,
i.e. $t_\perp/m= {\cal O}(1)$. In order to understand this result
it is instructive to consider the case
\be
T_\perp({\bf k})=t_\perp\sum_{j=1}^D\cos k_j\ ,\quad D\gg 1.
\ee
In this limit $T_\perp({\bf k})$ can be considered to be a random
variable with probability distribution
\be
P(t)=\frac{1}{\sqrt{\pi D
t_\perp^2}}\exp\left(-\frac{t^2}{Dt_\perp^2}\right). 
\label{poft}
\ee
As we are interested in the formation of a Fermi surface we have to
consider $T_\perp({\bf k})={\cal O}(m)$, which implies that $t_\perp
D/m={\cal O}(1)$. This leads to the restriction 
that for large $D$ 
\be
t_\perp\propto \frac{m}{D}\ ,
\ee
so that the probability distribution \r{poft} becomes extremely narrow
and only regions in k-space with $T_\perp({\bf k})\approx 0$
contribute to the DOS. However, in these regions there are no states
at $\omega\approx 0$ and there will therefore be no peak in the DOS at
low energies.

\section{Bechgaard salts as a possible application}
\label{sec:bechgaard}
Our theory may be relevant to the Bechgaard salts and in particular to
(TMTTF)$_2$PF$_6$ and (TMTSF)$_2$PF$_6$. (see \cite{Ishiguro} ,
\cite{review} for a review). However, for our theory
to be relevant, certain conditions must be met and this requires a
discussion. The materials in question are quasi-one-dimensional and
have a $3/4$-filled band. The ratio of the hopping integrals in the
three principal crystallographic directions is $t_a : t_b : t_c = 1 :
0.1 : 0.005$. Therefore, at sufficiently high temperatures one may
neglect the hopping in the $c$-direction. Then one is left with a
two-dimensional system of chains where each chain has only two nearest
neighbours (${\cal N} = 2$). Such small number of nearest neighbours
may put into question the applicability of RPA. This is, however, not
the main worry.

The principal problem one encounters in dealing with the Bechgaard
salts is the problem identifying the correct low-energy effective
theory. Let us neglect the interchain hopping for the time being and
consider uncoupled 1D chains. In the Bechgaard salts there are two
separate mechanisms that have the potential to open a spectral gap.
Firstly there are ``double'' Umklapp processes due to the commensurate
band filling 3/4 \cite{4umklapp,yoshioka}. These processes involve
the scattering of four electrons at $k_F$ and four holes at $-k_F$ and
are generated in the low-energy effective theory by integrating out
high energy modes. These processes will generate a gap only for strong
interactions ($K_c<0.25$). Secondly there is a small dimerization,
which halves the Brillouin zone and gives rise to ``single'' Umklapp
processes \cite{vic,penc2}. These open a gap already for weak
interactions $K_c<1$ but their coupling constant is proportional to
the dimerization and thus small. 
At low energies the system is thus described by two independent
Gaussian models - one for the charge and the other for the spin
sector. The charge sector Gaussian model is perturbed by two
operators: the $4k_F$-harmonics (with $k_F$ being equal to $3\pi/2a$)
of the dimerization 
\bea
\hat D = \delta t\sum_n (-1)^n  c^+_{n+1,\sigma}c_{n,\sigma}
\eea
where $\delta t$ is a staggered component of the hopping integral,
and the $8k_F$ component of the electron density operator
(double-Umklapp processes). Since these operators have different
symmetry under parity transformation (one is defined on links and the
other on sites), in the continuous limit they are given by sin and cos
respectively, such that the related Hamiltonian density is  
\bea
V = g_1 \sin(\sqrt{8\pi K_c}\Phi_c) + g_2\cos(2\sqrt{8\pi K_c}\Phi_c)\
.
\label{V}
\eea

The behaviour of the system and applicability of our theory depend
crucially on the value of $K_c$. We can consider the following
possibilities.

(i) The interactions are weak and $K_c \approx 1$ \cite{vic,braz}. That
the interactions are moderate is suggested by the renormalization of
the uniform magnetic susceptibility with respect to the value
extracted from the band structure calculations: $\chi_s/\chi_0 \approx
2-3$ \cite{review}. In this case, the $g_2$-term in (\ref{V}) is
irrelevant and the first term (due to dimerization) gives the
sine-Gordon model which is equivalent to the charge sector of the
Thirring model we have discussed. All our calculations are valid in
this case. 

If we adopt this scenario we have problems with Angle Resolved
Photoemission (ARPES) data which do not show any traces of
quasiparticles \cite{zwick,private}. We will return to the
question about ARPES data later.

(ii) There is a school of thought which advocates the small value
$K_c \approx 0.22$ when both operators are relevant \cite{schwarz}.
According to this school, the dimerization coupling is small and 
the $g_2$-term dominates. This would agree qualitatively  
with ARPES data. 

If this point of view is correct, the  calculations presented in this
paper are not applicable because, as it follows from \cite{LukZam01}, 
for such values of $K_c$ the minimal formfactor corresponds to the 
emission of not one, but two solitons. We have discussed  this
scenario in our other  publication \cite{quarter}. Here we cite just
one  conclusion from  this paper: if  $K_c$ is small and the
dimerization term is negligible, the value of the gap measured by
ARPES should be equal to the optical gap observed in the frequency
dependence of the on-chain conductivity, or twice the value of the
activation gap in the temperature dependence of the dc conductivity:
\bea
\Delta_{\rm ARPES} = 2E_{\rm act} = \Delta_{\rm opt} = 2m
\eea
Indeed, in (TMTTF)$_2$PF$_6$ ARPES measures a gap $\Delta \approx 100$
meV and thermal conductivity measurements give the activation gap
$E_{\rm act} = 44 meV$ \cite{Vescoli}, which is roughly one half.

(iii) The third possible scenario is that $K_c$ is small, but 
the dimerization is not negligible. To resolve this issue
theoretically one has to estimate the bare values of the coupling
constants $g_{1,2}$. We know how to do this only for small
$U$, where perturbation theory gives the following estimate
\[
g_2 \sim E_0(U/E_0)^3\ .
\]

If both terms are relevant and the sign of $g_2$ is positive, as it
follows from the perturbative calculation, then the $g_1$-term leads to
the confinement of the solitons in the $g_2$-sine-Gordon model. This
is a rather difficult case for the theory because the double
sine-Gordon model is not integrable.

Let us assume for the time being that $K_c \approx 1$ and see how our
theory would square with experiment. (TMTTF)$_2$PF$_6$ is a Mott
insulator; the Fermi energy is estimated as 115 meV,  
the hopping integral in the $b$-direction is of order of 14 meV, the
optical gap $\Delta_{\rm opt} = 2m$ is of order of 900K. Here the
transverse tunneling is not large enough to overcome the Mott gap. In
trying to get a detailed comparison with our calculations one has
to take into account that the ratio $m/\epsilon_F \approx 1/3$ is not
very small here, which reduces the chances of obtaining a quantitative
agreement with any field theoretical approach.  

(TMTSF)$_2$PF$_6$ is metallic; the Fermi energy is estimated as 220
meV, the hopping integral in the $b$-direction is of order of 20 meV,
the optical gap $\Delta_{\rm opt} = 2m$ is of order of 250K. This gives
$T_\perp(0)/m \approx 4$ so that the criterion for having a small
Fermi surface $Z_0T_\perp(0)/m > 3.3 $ is satisfied here.  
Optical measurements for this material show a metallic Drude peak with a 
tiny amount of spectral weight (3 percent), separated by a gap from a
very strong continuum. The metallic character of these compounds
is due to the transverse hopping \cite{Giam},\cite{Vescoli}. 
For frequencies not too close to the gap the observed form of the
optical conductivity in the direction along the chains is well
described by the sine-Gordon model \cite{CET}. This fact together with
the observed smallness of the Drude weight indicate that
(TMTSF)$_2$PF$_6$ is a good candidate for application of the present
theory.

An additional argument in favour of small pockets of Fermi liquid is
that not all physical properties of (TMTSF)$_2$PF$_6$ demonstrate the
same degree of anisotropy. For example, the measured anisotropy of
the plasma frequency for the Drude peaks is only of a factor of
2\cite{Henderson}. On the other hand, for the ratio of  hopping
integrals predicted by the band  theory one should expect it to be of
order of $(m_{\perp}/m_{\parallel})^{1/2} \approx  10$. 
This fact in combination with the smallness of the Drude weight indicates
that the Fermi surface is small and not very anisotropic. On the other
hand, many of the properties of these material (especially the
magnetic ones) are typically one-dimensional 
(see \cite{review} for review). Thus the overall picture is in
reasonable agreement with the scenario we present. 

The analysis of the Drude peak given in \cite{schwarz} indicates that
the best fit can be obtained if one assumes frequency dependent
effective mass and the scattering rates in the Drude formula: 
\bea
\sigma(\omega) = \frac{\omega_p^2}{4\pi}\frac{1}{\Gamma_1(\omega) -
\ri\omega[\frac{m^*(\omega)}{m_{\rm band}}]},\nonumber\\ 
\frac{m^*(\omega)}{m_{\rm band}} = 1 + \frac{\lambda_0}{1 + \alpha^2\omega^2}, \nonumber\\
\Gamma_1(\omega) = \Gamma_0 + \frac{\lambda_0\alpha\omega^2}{1 +
\alpha^2\omega^2} \ .
\eea
This fit is rather suggestive because the frequency dependence is
quadratic, like in Fermi liquid theory. This feature supports the
point of view that the Drude peak comes from small pockets of Fermi liquid.

There are other quasi-one-dimensional systems for which ARPES
measurements have been performed:
the blue bronze K$_{0.3}$MoO$_3$, which is metallic \cite{exp1}, and the
Mott insulator Sr$_2$CuO$_3$. The latter material, however, is not a
good testing ground for our theory since the ratio of the Mott-Hubbard gap
to the bandwidth is too large. The largeness of the gap precludes a
detailed comparison with the results obtained in this paper. On the
qualitative side the measurements demonstrate the appearence of two distinct 
dispersing maxima in the spectral function \cite{exp}, which is
interpreted as a sign of spin-charge separation.

ARPES is not the only way to ascertain the existence of
quasi-particles. de Haas-van Alphen and Schubnikov-de Haas effects
are perfect tools when one deals with closed Fermi
surfaces. Obviously, the measurements should be made above the
ordering temperature which may impose serious difficulties.  

\acknowledgements
This work was started at the Isaac Newton Institute for Mathematical
Sciences in Cambridge during the program on Strongly Correlated
Electrons in 2000. We are grateful to the Institute for
hospitality. Our warmest thanks are to Sergei Lukyanov for his help
and interest to the work. We acknowledge illuminating conversations
with A. Chubukov, N. D'Ambrumenil, A. Finkelstein, T. Giamarchi,
A. Georges, V. Yakovenko  and P. Wiegmann. We also thank S. Kivelson
and D. Orgad for useful correspondence. This work was supported by the
EPSRC under grants AF/100201 and GR/N19359 (FHLE) and by the DOE under
contract number DE-AC02-98 CH 10886. 

\appendix
\section{Elements of RPA}
\label{sec:RPA}
\begin{figure}[ht]
\begin{center}
\epsfxsize=0.4\textwidth
\epsfbox{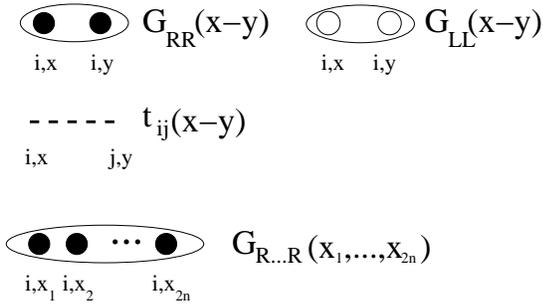}
\end{center}
\caption{\label{fig:rpa1}
Elements of the diagram technique in position space.
Only one kind of many-particle ``vertex'' is shown. 
}
\end{figure}

\begin{figure}[ht]
\begin{center}
\epsfxsize=0.35\textwidth
\epsfbox{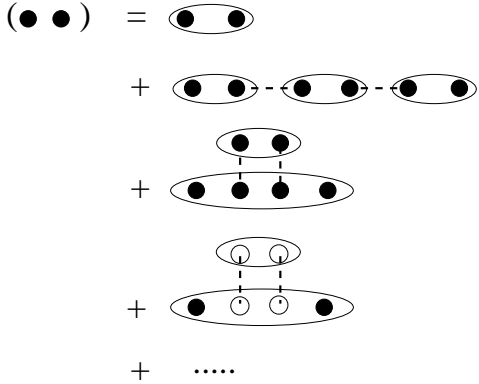}
\end{center}
\caption{\label{fig:diag2a}
Diagrams for initial and final points on the same chain up to second
order in the interchain hopping. 
}
\end{figure}

In this appendix we discuss some relevant aspects of the RPA in the
interchain coupling. This is most easily done in position space.
We denote the right and left-moving fermion operators by black and
white circles. 1D correlation functions are denoted by encircling a
number of circles, the corresponding fermion operators are then all
located on the same chain. In Fig.\ref{fig:rpa1} we show the
corresponding diagrams for the 1D two-point functions of right movers
and left movers as well as the diagram fro the 2n-point function of
right movers. Finally, we denote the interchain hopping matrix
element $t_{ij}(x-y)$ between sites $x/a_0$ on chain $i$ and site
$y/a_0$ on chain $j\neq i$ by a dashed line. We note that the hopping
is local in time. 
The first few diagrams in the expansion (in the interchain hopping) of
the two-point function of right moving electrons for initial and final
point located on the same chain is shown in Fig.\ref{fig:diag2a}. The
contribution of a given diagram is obtained by summing over the
positions of all ``internal'' circles, that is circles connected by a
hopping line. 

\begin{figure}[ht]
\begin{center}
\epsfxsize=0.4\textwidth
\epsfbox{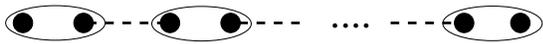}
\end{center}
\caption{\label{fig:rpa2}
Diagrams extering the RPA expression for the single-particle
Green's function. The sum is over all diagrams of the type shown.
}
\end{figure}

\begin{figure}[ht]
\begin{center}
\epsfxsize=0.25\textwidth
\epsfbox{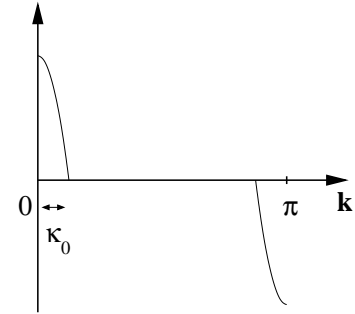}
\end{center}
\caption{\label{fig:ttilde}
Schematic dependence of $T_\perp({\bf{k}})$ on the transverse momenta
$q_\perp$. 
}
\end{figure}

The RPA expression for the (Fourier transform) of the single-particle
Green's function is obtained by summing all diagrams that can be split
into two parts by cutting any one hopping line, i.e. all diagrams of
the type shown in Fig.\ref{fig:rpa2}.
All diagrams neglected in RPA all contain loops. This enables us
to embed RPA into a systematic perturbative expansion in a small
parameter $\kappa_0$ as follows. Let us consider an interchain
hopping $T_\perp({\bf k})$ (for simplicity we take it independent of
the wave number $q$ along the  chain direction) of the form shown in
Fig.\ref{fig:ttilde} i.e. particle-hole symmetric but long ranged such
that its Fourier transform is strongly peaked around the origin and
the point $(0,\pi,\pi)$. This means that in position space the hopping
is {\sl long-ranged}, i.e. the hopping amplitudes are of the same
order within a range proportional to $\kappa_0^{-1}$.
Every loop gives a contribution
\be
\frac{1}{V}\sum_{{\bf k}}\left[T_\perp({\bf k})\right]^2\propto 
\kappa_0^2
\ee
and is thus suppressed. In this way one obtains a formal expansion in
powers of $\kappa_0^2$, the leading ($\kappa^0$) order of which is
given by the RPA.

\section{Formfactor Approach in integrable Quantum Field Theories}
\label{sec:FF}
The spectrum of low-lying excitations of the half-filled Hubbard model
consists of scattering states of gapped, spinless charge $\pm e$
excitations called {\sl holons} and {\sl  antiholons} and
gapless, charge-neutral excitations carrying spin $\pm\frac{1}{2}$,
the so-called {\sl spinons}\cite{smat}. We introduce labels
$h$,$\bar{h}$,$s$,$\bar{s}$ to distinguish between these four types of
elementary excitations. Their dispersions and exact
scattering matrices are known on the lattice \cite{smat} as well as in
the field theory limit \cite{ezer}.
As usual for particles with relativistic dispersion it is useful to
introduce rapidity variables $\theta_{c,s}$ to parametrize energy and
momentum 
\begin{eqnarray}
E_{\alpha}(\theta_c)&=&m\cosh\theta_c\
,P_{\alpha}(\theta_c)=m\sinh\theta_c\ ,\ \alpha=h,\bar{h},\nn
E_{\gamma}(\theta_s)&=&\frac{m}{2}e^{\pm\theta_s}\
,P_{\gamma}(\theta_s)=\pm\frac{m}{2}e^{\pm \theta_s}\ ,\gamma=s,\bar{s}.
\end{eqnarray}
Here we have set spin and charge velocities to $1$. Let us now
turn to the constuction of a basis of scattering states of holons,
antiholons and spinons. A convenient formalism to this end is 
is obtained in terms of the Zamolodchikov-Faddeev (ZF) algebra. The ZF
algebra can be considered to be the extension of the algebra
of creation and annihilation operators for free fermion or bosons to
the case or interacting particles with factorizable scattering.
The ZF algebra is based on the knowledge of the exact spectrum and
scattering matrix \cite{smat}. For the SGM
the ZF operators (and their hermitian
conjugates) satisfy the following algebra
\begin{eqnarray}
{Z}^{\epsilon_1}(\theta_1){Z}^{\epsilon_2}(\theta_2) &=&
S^{\epsilon_1,\epsilon_2}_{\epsilon_1',\epsilon_2'}(\theta_1 -
\theta_2){Z}^{\epsilon_2'}(\theta_2){Z}^{\epsilon_1'}(\theta_1)\ ,
\nonumber\\
{Z}_{\epsilon_1}^\dagger(\theta_1)Z_{\epsilon_2}^\dagger(\theta_2) &=&
Z_{\epsilon_2'}^\dagger(\theta_2){ Z}_{\epsilon_1'}^\dagger
(\theta_1)S_{\epsilon_1,\epsilon_2}^{\epsilon_1',\epsilon_2'}(\theta_1 -
\theta_2) , \nonumber\\
Z^{\epsilon_1}(\theta_1)Z_{\epsilon_2}^\dagger(\theta_2) &=&Z_{\epsilon_2'}
^\dagger(\theta_2)
S_{\epsilon_2,\epsilon_1'}^{\epsilon_2',\epsilon_1}
(\theta_2-\theta_1)Z^{\epsilon_1'}(\theta_1)\nonumber\\
&&+(2 \pi) \delta_{\epsilon_2}^{\epsilon_1} 
\delta (\theta_1-\theta_2).
\label{fz1}
\end{eqnarray}

Here $S^{\epsilon_1,\epsilon_2}_{\epsilon_1',\epsilon_2'}(\theta)$ are
the (factorizable) two-particle scattering matrices and
$\varepsilon_j=s,\bar{s},h,\bar{h}$.

Using the ZF generators a Fock space of states can be constructed as
follows. The vacuum is defined by
\begin{equation}
Z_{\varepsilon_i}(\theta) |0\rangle=0 \ .
\end{equation}
Multiparticle states are then obtained by acting with strings of
creation operators $Z_\epsilon^\dagger(\theta)$ on the vacuum
\begin{equation}
|\theta_n\ldots\theta_1\rangle_{\epsilon_n\ldots\epsilon_1} = 
Z^\dagger_{\epsilon_n}(\theta_n)\ldots
Z^\dagger_{\epsilon_1}(\theta_1)|0\rangle . 
\label{states}
\end{equation} 
In term of this basis the resolution of the identity is given by
\begin{equation}
\sum_{n=0}^\infty\sum_{\epsilon_i}\int_{-\infty}^{\infty}
\frac{d\theta_1\ldots d\theta_n}{(2\pi)^nn!}
|\theta_n\ldots\theta_1\rangle_{\epsilon_n\ldots\epsilon_1}
{}^{\epsilon_1\ldots\epsilon_n}\langle\theta_1\ldots\theta_n|\ .
\label{identity}
\end{equation}
Inserting (\ref{identity}) between operators in a 2-point function
we obtain the following spectral representation 

\begin{eqnarray}
\label{corr}
&&\langle {\cal O}(x,t){\cal O}^\dagger(0,0)\rangle
=\sum_{n=0}^\infty\sum_{\epsilon_i}\int
\frac{d\theta_1\ldots d\theta_n}{(2\pi)^nn!}
\nonumber\\
&&\qquad\times\
\exp\Bigl({i\sum_{j=1}^n P_{\epsilon_j}(\theta_j)x-
E_{\epsilon_j}(\theta_j) t}\Bigr)\nn
&&\qquad\times\ 
|\langle 0| {\cal O}(0,0)|\theta_n\ldots\theta_1
\rangle_{\epsilon_n\ldots\epsilon_1}|^2, 
\end{eqnarray}
where
\begin{equation}
\label{formf}
f^{\cal O}(\theta_1\ldots\theta_n)_{\epsilon_1\ldots\epsilon_n}\equiv
\langle 0| {\cal
O}(0,0)|\theta_n\ldots\theta_1\rangle_{\epsilon_n\ldots\epsilon_1} 
\end{equation}
are the form factors.


\begin{thebibliography}{99}
\bibitem{comment1}
Here we consider transverse hopping between n.n. chains,
so that RPA is not a controlled approximation.
\end{thebibliography}
\end{document}